\documentclass[fleqn,usenatbib]{mnras}

\usepackage{newtxtext,newtxmath}
\usepackage{graphicx}
\usepackage{bm}
\usepackage{cancel}
\usepackage{epsfig}
\usepackage{dcolumn}
\usepackage{amsmath}
\usepackage{hhline}
\usepackage{enumerate}
\usepackage[utf8]{inputenc}
\usepackage[dvipsnames]{xcolor}
\usepackage{hyperref}
\usepackage{mwe}
\usepackage{widetext}
\usepackage{wasysym}
\usepackage{longtable}
\usepackage{pdflscape}
\usepackage{tikz} 
\usepackage[T1]{fontenc}

\DeclareRobustCommand{\VAN}[3]{#2}
\let\VANthebibliography\thebibliography
\def\thebibliography{\DeclareRobustCommand{\VAN}[3]{##3}\VANthebibliography}

\def \beq  {\begin{equation}}
\def \eeq  {\end{equation}}
\def \ber  {\begin{eqnarray}}
\def \eer  {\end{eqnarray}}

\def \om    {\Omega}

\def \om0m {\Omega_{0\rm m}}

\title{Probing dark fluids and modified gravity with gravitational lensing}

\author[Leandros Perivolaropoulos, Ioannis Antoniou \& Demetrios Papadopoulos]{
Leandros Perivolaropoulos \orcidA{},$^{1}$\thanks{Contact e-mail: \href{mailto:leandros@uoi.gr}{leandros@uoi.gr}}%
Ioannis Antoniou \orcidB{},$^{1}$\thanks{Contact e-mail: \href{mailto:i.antoniou@uoi.gr}{i.antoniou@uoi.gr}}%
Demetrios Papadopoulos, $^{2}$\thanks{Contact e-mail: \href{mailto:papadop@astro.auth.gr}{papadop@astro.auth.gr}}
\\
$^{1}$Department of Physics, University of Ioannina, GR-45110, Ioannina, Greece\\
$^{2}$Department of Physics, Aristotle University of Thessaloniki, Section of Astrophysics, Astronomy and Mechanics, 54124 Thessaloniki, Greece
}

\pubyear{2023}

\begin{document}

\interfootnotelinepenalty=10000

\newcommand{\newc}{\newcommand}

\newcommand{\orcidauthorA}{0000-0001-9330-2371} 
\newcommand{\orcidauthorB}{0000-0002-3477-7876} 

\newcommand{\be}{\begin{equation}}
\newcommand{\ee}{\end{equation}}
\newcommand{\ba}{\begin{eqnarray}}
\newcommand{\ea}{\end{eqnarray}}
\newcommand{\bea}{\begin{eqnarray*}}
\newcommand{\eea}{\end{eqnarray*}}
\newc{\D}{\partial}
\newc{\ie}{{\it i.e.} }
\newc{\eg}{{\it e.g.} }
\newc{\etc}{{\it etc.} }
\newc{\etal}{{\it et al.}}
\newc{\lcdm}{$\Lambda$CDM }
\newc{\lcdmnospace}{$\Lambda$CDM}
\newc{\wcdm}{$w$CDM }
\newc{\plcdm}{Planck18/$\Lambda$CDM }
\newc{\plcdmnospace}{Planck18/$\Lambda$CDM}
\newc{\omom}{$\Omega_{0m}$ }
\newc{\omomnospace}{$\Omega_{0m}$}
\newcommand{\nn}{\nonumber}
\newc{\ra}{\Rightarrow}
\newc{\baodv}{$\frac{D_V}{r_s}$ }
\newc{\baodvnospace}{$\frac{D_V}{r_s}$}
\newc{\baoda}{$\frac{D_A}{r_s}$ } 
\newc{\baodanospace}{$\frac{D_A}{r_s}$}
\newc{\baodh}{$\frac{D_H}{r_s}$ }
\newc{\baodhnospace}{$\frac{D_H}{r_s}$}

\newcommand{\orcidicon}{\includegraphics[width=0.32cm]{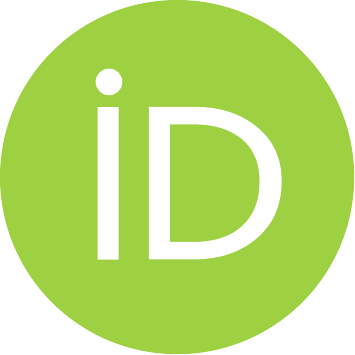}}

\foreach \x in {A, ..., Z}{%
\expandafter\xdef\csname orcid\x\endcsname{\noexpand\href{https://orcid.org/\csname orcidauthor\x\endcsname}{\noexpand\orcidicon}}
}

\label{firstpage}
\pagerange{\pageref{firstpage}--\pageref{lastpage}}
\maketitle

\begin{abstract}
We generalize the Rindler-Ishak (2007) result for the lensing deflection angle in a SdS spacetime,  to the case of a general spherically symmetric fluid beyond the cosmological constant. We thus derive an analytic expression to first post-Newtonian order for the lensing deflection angle in a general static spherically symmetric metric of the form $ ds^2 = f(r)dt^{2} -\frac{dr^{2}}{f(r)}-r^{2}(d\theta ^2 +\sin ^2 \theta d\phi ^2)$ with $f(r) = 1 - \frac{2m}{r}-\sum_{i}  b_i\; r_0^{-q_i}\; \left( \frac{r_0}{r}\right)^{q_i}$ where $r_0$ is the lensing impact parameter, $b_i\ll r_0^{q_i}$, $m$ is the mass of the lens and $q_i$ are real arbitrary constants related to the properties of the fluid that surrounds the lens or to modified gravity. This is a generalization of the well known Kiselev black hole metric. The approximate analytic expression of the deflection angle is verified by an exact numerical derivation and in special cases it reduces to results of previous studies.  The density and pressure of the spherically symmetric fluid that induces this metric is derived in terms of the constants $b_i$. The Kiselev case of a Schwarzschild metric perturbed by a general spherically symmetric dark fluid (eg vacuum energy) is studied in some detail and consistency with the special case of Rindler Ishak result is found for the case of a cosmological constant background. Observational data of the Einstein radii from distant clusters of galaxies lead to observational constraints on the constants $b_i$ and through them on the density and pressure of dark fluids, field theories or modified gravity theories that could induce this metric.
\end{abstract}




\section{Introduction}
Cosmological observations have indicated that about $95\%$ of the energy content of the universe is of unknown origin. About $25\%$ of this unknown energy, known as {\it dark matter \cite{Bertone:2016nfn} } behaves like a perfect fluid with equation of state 
$w\simeq 0$ which is similar to that of matter with velocity much smaller than the velocity of light $c$ that interacts only gravitationally. The other $70\%$ usually called {\it dark energy \cite{Frieman:2008sn, Copeland:2006wr}} behaves like a perfect fluid with equation of state $w\simeq -1$ which is similar to that of a cosmological constant $\Lambda$ \cite{Padmanabhan:2002ji, Peebles:2002gy, Carroll:2000fy, Sahni:1999gb}. The {\it standard $\Lambda CDM$ model \cite{Planck:2018vyg} } assumption is that dark matter consists of a particle which can be discovered in accelerator experiments while dark energy is actually the cosmological constant. This interpretation however is being challenged by three facts: 
\begin{itemize}
    \item Despite long and persistent efforts of a few decades it has not been possible to identify the dark matter particle in Earth bound experiments \cite{Rogers:2020ltq,XENON:2019gfn}.
    \item The required cosmological constant  value is too low to be consistent with any particle physics theory (the fine tuning problem) \cite{Padilla:2015aaa}.
    \item The internal observational consistency of $\Lambda CDM$ has been challenged recently by conflicting best fit values of parameters (tensions \cite{Perivolaropoulos:2021jda}) of the standard model. The most prominent and persistent of these tensions is the {\it Hubble tension \cite{DiValentino:2021izs}}: The Hubble parameter $H_0$ as measured from the CMB sound horizon \cite{Planck:2018vyg} standard ruler under the assumption of $\Lambda CDM$ is in $5\sigma$ conflict with the best fit value obtained using the local distance ladder method with Type Ia Supernovae (SnIa) \cite{Riess:2021jrx}.
\end{itemize}

It is therefore becoming increasingly likely that the assumed properties of the two main fluids of the universe may deviate from the standard model assumptions. 

One of the most efficient probes of the detailed properties of cosmological fluids is gravitational lensing \cite{Zwicky:1937zzb,Dyson:1920cwa,Walsh:1979nx,Bozza:2010xqn,Bartelmann:2010fz,Cunha:2018acu,He:2017alg,Piattella:2016nzt,Ali:2017ofu,Lake:2001fj,Rindler:2007zz,Takizawa:2021jxa,Virbhadra:2008ws,Virbhadra:1999nm,Wambsganss:1998gg}. Gravitational lensing can probe directly the local metric parameters in a generic model independent manner and therefore is a useful tool for the detection of signatures of either exotic fluids \cite{Finelli:2006iz} or modified gravity \cite{Mannheim:2005bfa,Wheeler:2013ora,Kiefer:2017nmo,Li:2011ur}. In the presence of such effects the General Relativistic (GR) vacuum metric would get modified \cite{Mannheim:1988dj,Edery:1997hu,Cutajar:2014gfa,Grumiller:2010bz} at both the solar system \cite{Ozer:2017oik,Edery:1997hu,Kagramanova:2006ax,Sereno:2007rm,Iorio:2007ub} and the galactic and cluster scales \cite{Varieschi:2008va,Chang:2011bp,Pizzuti:2017diz}. 
Such modifications would need to be distinguished from other effects like non-spherically symmetric matter near a gravitational lens galaxy/cluster or projected along the line of sight \cite{McCully:2016yfe}. Despite of these effects, upper bounds on the spherical metric parameters can still be obtained by assuming that any deviation from the Schwarzschild metric is due to dark fluids and not to other effects. In this context, any order of magnitude estimate of the extended metric parameters would be considered as an upper bound.

The deflection angle of the light emitted by a background source deflected by a foreground lens (eg cluster) is a quantity \cite{DES:2017tby} that depends on the metric parameters \cite{Lim:2016lqv}.

Even though such strong lensing systems are difficult to identify \cite{Metcalf:2018elz,Jacobs:2017xhn,Petrillo:2017njm}, in the context of an approximately spherically symmetric lens, the measured deflection angle can lead to direct measurement of the metric parameters provided that the metric is modeled in a general enough context \cite{Rindler:2007zz,Jha:2023qqz,Ishak:2007ea,Ishak:2008zc,Sultana:2012zz,He:2017alg,Azreg-Ainou:2017obt,Younas:2015sva,Lim:2016lqv, Kitamura:2012zy}. The simplest modeling of the metric around a lens system is the Schwarzschild vacuum metric which has been extensively used for the search of unseen matter associated with the lens. In the context of this metric, the deflection angle to lowest order is \cite{Weinberg:1972kfs} (In what follows we set Newton's constant $G$ and the speed of light $c$ to unity unless otherwise mentioned.)
\begin{equation} 
{\hat \alpha}=\frac{4m}{r_0}
\label{hatasch1}
\end{equation}
where $m$ is the mass of the lens and $r_0$ is the distance of closest approach of the light-ray to the lens (the impact parameter). Thus, measurement of $\hat \alpha $ can lead to estimates and constraints on the mass $m$ of the lens and comparison with the visible part of the mass can lead to estimate of the dark matter content of the lens. 

In the context of generalized metrics, the additional parameters may also be constrained by the measurement of $\hat \alpha$. For example in the presence of vacuum energy (a cosmological constant $\Lambda$ with a term $\frac{\Lambda}{3} r^2$ in the spherically symmetric metric) the predicted deflection angle becomes \cite{Rindler:2007zz} 
\begin{equation}
{\hat \alpha_{SdS}}\simeq \frac{4m}{r_0} (1-\frac{2m^2}{r_0^2}-\frac{\Lambda r_0^4}{24m^2})
\label{hatasds0}
\end{equation}
Using cluster lensing data,  this form of generalized $\hat \alpha$ has led to constraints of the value of $\Lambda$ ($\Lambda \lesssim 10^{-54} cm^{-2}$ \cite{Ishak:2007ea}) which approaches the precision of corresponding cosmological constraints obtained from measurements of the expansion rate of the Universe at various redshifts $\Lambda \simeq 10^{-56} cm^{-2}$. Similarly a generalized spherically symmetric metric with a Rindler term $\sim b\; r$ has led to constraints on the Rindler acceleration term $b\lesssim 10^{-2} m/sec^2$ from solar system quasar lensing data \cite{Carloni:2011ha}. It is therefore interesting to consider other spherically symmetric generalizations of the Schwarzschild metric and impose constraints on their parameters using gravitational lensing data. Previous studies have investigated the effects of special cases of spherical generalized metrics \cite{Zhang:2021ygh} on gravitational lensing \cite{Azreg-Ainou:2017obt,Younas:2015sva} and other observables \cite{Sheykhi:2010yq} like galactic clustering \cite{Khanday:2021kjy}. 

In the present analysis, we derive general analytical expressions of the predicted deflection angle in the context of a wide range of spherically symmetric metrics. This class of metrics includes as special cases the Schwarzschild deSitter (SdS) metric \cite{Rindler:2007zz}, the Reissner-Nordstrom metric \cite{Eiroa:2002mk}, nonlinear electrodynamics charged black hole \cite{Gurtug:2020wpi,Gurtug:2019yzu}, Yukawa black holes \cite{Benisty:2022txp}, the global monopole metric \cite{Barriola:1989hx, Platis:2014sna},  the Rindler-Grumiller metric \cite{Carloni:2011ha,Grumiller:2010bz,Perivolaropoulos:2019vgl,Lim:2019akv,Sultana:2012qp,Gregoris:2021plc}, the Weyl gravity vacuum metric \cite{Mannheim:1988dj,Edery:1997hu} the $SdS_w$ metric \cite{Zhang:2021ygh,Fernando:2012ue,Uniyal:2014paa} the Kiselev black hole \cite{Kiselev:2002dx,Younas:2015sva,Liu:2022hbp,Alfaia:2021cnk,Shchigolev:2016gro,Abbas:2019olp}, the Kiselev charged black hole \cite{Azreg-Ainou:2017obt,Atamurotov:2022knb} and the interior Kottler metric \cite{Schucker:2010rkd, Antoniou:2016obw} (for a good review of such spherical inhomogenous solutions see \cite{Faraoni:2021nhi}). Then we compare these expressions with the measured values of the deflection angle in the context of cluster scale systems thus imposing constraints on the metric parameters that appear in the analytic expressions of the deflection angle $\hat \alpha$. In this context, after deriving the analytical expressions for $\hat \alpha$, we use  observations of Einstein radii around distant galaxies and clusters of galaxies to derive the measured lensing deflection angle. Using observational data of a selected list
of Einstein radii around clusters and galaxies, we derive upper bound, order of magnitude constraints, on the new metric parameters in the context of a wide range of models.  These results provide an
improvement of several orders of magnitude on previous upper bounds on these parameters from planetary or stellar systems \cite{Sereno:2007rm,Kagramanova:2006ax}.

The structure of this paper is the following: In the next section we consider a generalized spherically symmetric metric and connect its parameters with a possible exotic fluid energy-momentum tensor that could give rise to it. In section \ref{III} we derive general analytic expressions for the deflection angle of such metrics. In section \ref{IV} we apply these analytic expressions to derive the deflection angle in a Schwarzschild metric perturbed by a general power-law term (Kiselev metric) which may represent either an exotic fluid or a modification of GR in the vacuum. Special cases of such a term include a vacuum energy term, a Rindler acceleration term, a global monopole scalar field gravity, the electric field of Reissner-Nordstrom metric or other more general terms. In section \ref{V} we compare the deflection angle of the perturbed Schwarzschild metric of section \ref{IV} with the measured Einstein radii and deflection angles around clusters and derive order of magnitude constraints on the new metric parameters of the perturbing power-law terms. Finally in section \ref{VI}, we conclude, summarize and discuss possible future prospects of the present analysis.

\section{General Class of Spherically Symmetric Metrics and their fluid background}
\label{II}

We focus on the following class of spherically symmetric metrics 
\be
ds^2= f(r) dt^2 - f(r)^{-1} dr^2 - r^2 (d\theta^2 +\sin^2\theta d\phi^2)
\label{sphmetric}
\ee
The energy momentum tensor that can give rise to this metric may be obtained from the Einstein tensor $G_\nu^\mu$. For 
\be 
f(r)=1-\frac{2m}{r} - g(r)
\label{fgen1}
\ee
where $g(r)$ is an arbitrary function, it is easy to show that
\begin{widetext}
\ba
G^{\mu}_{\nu}=
  \begin{bmatrix}
    \frac{g(r)}{r^2}+\frac{g'(r)}{r} & 0 & 0 & 0 \\
    0 & \frac{g(r)}{r^2}+\frac{g'(r)}{r} & 0 & 0 \\
    0 & 0 & \frac{g'(r)}{r}+\frac{g''(r)}{2} & 0 \\
    0 & 0 & 0 & \frac{g'(r)}{r}+\frac{g''(r)}{2}
  \end{bmatrix}
  =\kappa T^{\mu}_{\nu}
\ea
\end{widetext}
where $T^{\mu}_{\nu}$ is the energy momentum tensor that gives rise to the metric \eqref{fgen1}, \eqref{sphmetric} and $\kappa=8\pi G$. Clearly, the parameter $m$ does not appear in  $T^{\mu}_{\nu}$ because it corresponds to the vacuum solution. If $g(r)$ is a superposition of power law terms 
\be g(r)=\sum_i b_i r^{-q_i}
\label{grser}
\ee
where $q_i$ are arbitrary real constants, then the energy momentum of the fluid that supports the above metric may be written as \citep{Alestas:2019wtw}

\ba
T^{\mu}_{\nu}&=&\frac{1}{\kappa}\sum_{i} b_i (1-q_i) r^{-(q_i+2)}
  \begin{bmatrix}
    1 & 0 & 0 & 0 \\
    0 & 1 & 0 & 0 \\
    0 & 0 & -\frac{1}{2}q_i  & 0 \\
    0 & 0 & 0 &  -\frac{1}{2}q_i
  \end{bmatrix}\nonumber\\&=&
  \begin{bmatrix}
    \rho & 0 & 0 & 0 \\
    0 & -p_r & 0 & 0 \\
    0 & 0 & -p_\theta  & 0 \\
    0 & 0 & 0 &  -p_\phi
  \end{bmatrix}
  \label{tmunu}
\ea
where in the last equation we have denoted the fluid density $\rho$, the radial pressure $p_r$ and the tangential pressures $p_\theta=p_\phi$. This is a generalization of the well known Kiselev black hole \cite{Kiselev:2002dx}. Notice that since the fluid tangential and radial pressures are not equal, this energy momentum tensor does not correspond to a perfect fluid due to the inhomogeneity and local anisotropy of the pressure. Thus it is not directly related to quintessence as has been stated in previous studies. This issue is clarified in detail in Ref. \cite{Visser:2019brz}. However, in the combined presence of the terms $q_1=1$ (point mass), $q_2\simeq -2$ (cosmological constant) and $q_i>-2$ (dark fluids), the energy momentum tensor \eqref{tmunu} is consistent with dark energy because at large distances the cosmological constant $q_2\simeq -2$ term of the metric function $f(r)$ dominates and corresponds to a homogeneous and isotropic fluid with equation of state $w\simeq -1$. 

As expected, the term $q_i=1$ corresponds to zero energy momentum term (vacuum solution) while for $q_i=-2$ we obtain the cosmological constant term (constant energy density-pressure) and for $q_i=0$ we have the case of a global monopole \cite{Barriola:1989hx} (zero angular pressure components while the energy density, radial pressure drop as $\sim r^{-2}$). The electric field of a Reissner-Nordstrom black hole \citep{1916AnP...355..106R, 1918KNAB...20.1238N}  and the Ellis wormhole\cite{Ellis:1973yv} correspond to $q_i=2$  \cite{Abe:2010ap}. Other similar solitonic field configurations corresponding to different values of $q$ could in principle be constructed. A generic discussion of the sources generating metrics of this class of models can be found in \cite{Bozza:2015haa} (Section 2).
 
The question that we address in this analysis is the following: {\it 'What would be the signature of such general and exotic fluids in gravitational lensing?'} Such fluids would give rise to the metric \eqref{sphmetric} in the context of GR. A similar metric may also emerge in the context of modified GR gravity theories even as a vacuum solution \cite{Grumiller:2010bz,Ren:2021uqb,Shaikh:2017zfl}. Therefore the detection of signatures of such a generalized spherically symmetric metric could be interpreted either as a signature of an exotic fluid in the context of GR or as presence of modifications of GR. This prospect is investigated in the following sections.

\section{Photon Geodesics and Lensing in a general fluid}
\label{III}

\begin{figure}
\centering
\includegraphics[width = \columnwidth]{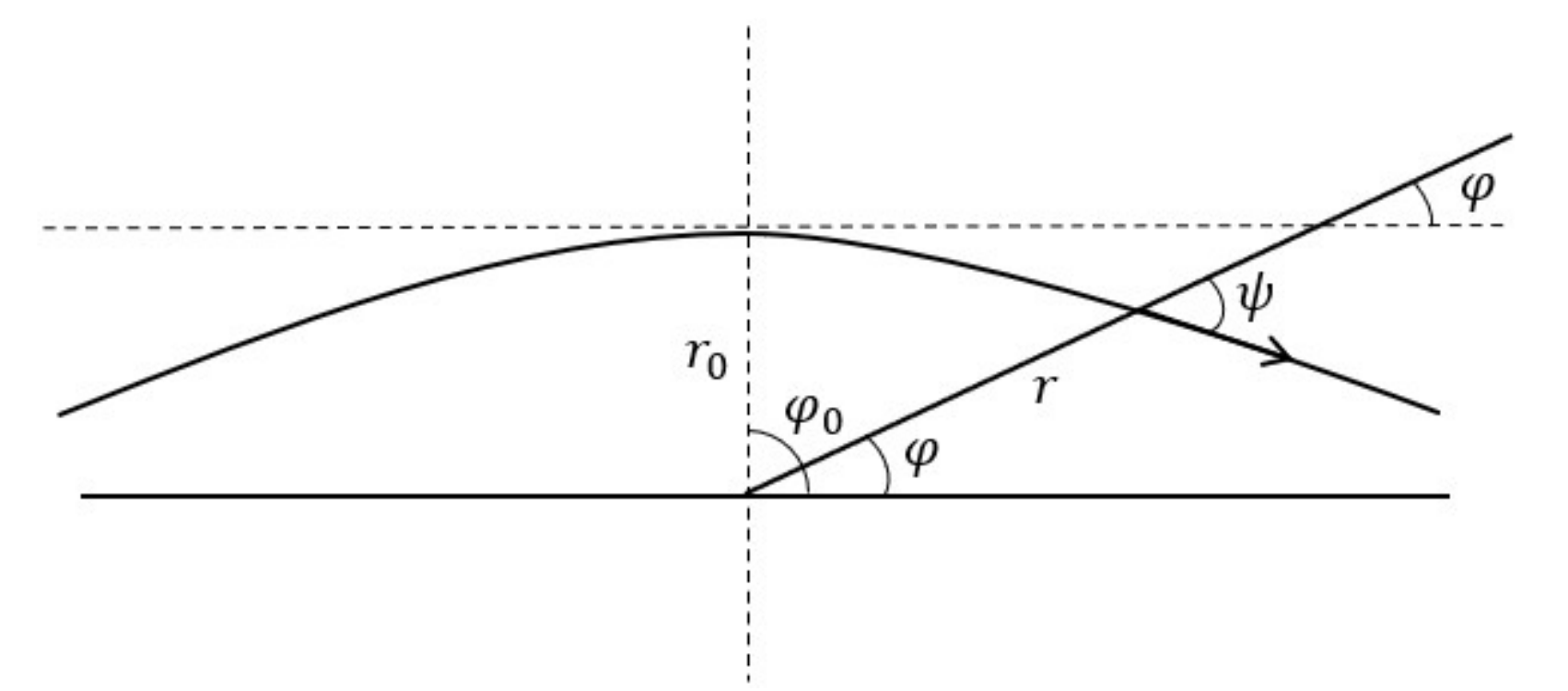}
\caption{The trajectory of light from source (left side) to observer (right side), passing at a distance of closest approach $r_0$ to the lens. We assume the trajectory does not cross either the cosmological or event horizons of the spacetime. Here we have drawn the angles $\phi_0$ and $\phi\equiv\phi_{obs}$ to be relative to the horizontal dashed line, implying that this horizontal line is the $\phi=0$
angle.}
\label{fig:lens}
\end{figure}

The geodesic equations are derived with respect to the Lagrangian $L=\frac{1}{2}g_{\mu\nu}\dot{x}^{\mu}\dot{x}^{\nu}$, where $\dot{x}^{\mu}(\lambda)$ is the time-like or null particle's trajectory and the dots denote derivatives with respect to the photon geodesic affine parameter $\lambda$. The Euler-Lagrange equation $\frac{d}{d\lambda}\frac{\partial L}{\partial\dot{x}^{\mu}}=\frac{\partial L}{\partial x^{\mu}}$ leads to the equations of motion
\begin{equation}
\dot{t} = \frac{E}{f} \label{tgeod}
\end{equation}
\begin{equation}
\dot{\phi} = \frac{h}{r^2 sin^{2}\theta} \label{phigeod}
\end{equation}

\begin{equation}
\ddot{r} = \frac{f'}{2f^2}\dot{r}^2 + rf\dot{\theta}^2 -\frac{f'E^2}{2f} + \frac{f\; h ^2}{r^3 sin^{2}\theta}
\end{equation}
\begin{equation}
\ddot{\theta} = -\frac{2\dot{r}\dot{\theta}}{r} + \frac{cos\theta \; h^{2}}{r^{4} sin^{3}\theta}
\end{equation}
where $E$, $h$ are the energy and the angular momentum of the particle respectively. The prime denotes derivative with respect to $r$. Use of these equations in the photon geodesic constraint
\begin{equation}
ds^2=g_{\mu\nu}\dot{x}^{\mu}\dot{x}^{\nu}=0
\label{photgeod}
\end{equation}
leads to
\begin{equation}
\dot{r}^2 + r^2 f\dot{\theta}^2 = E^2 -\frac{h^2 \; f}{r^2 sin^2\theta} 
\label{rdot}
\end{equation}
Fixing $\theta = \pi/2$, by spherical symmetry in (\ref{rdot}) leads to the radial equation
\begin{equation}
\dot{r}^2 = E^2 - \frac{h^2}{r^2}\;f(r)
\label{rdot1}
\end{equation}
At the distance of closest approach $r_0$ (see Fig. \ref{fig:lens}) where $\dot{r}=0$ we have
\begin{equation}
\frac{f(r_0)}{r_0^2}=\frac{E^2}{h^2}
\label{r0def}
\end{equation}
From eqs. (\ref{phigeod}), (\ref{rdot1}) and (\ref{r0def}) for $\theta=\pi/2$ we obtain the null geodesic trajectory equation
\begin{equation}
\left(\frac{dr}{d\phi}\right)^2 = r^4\left(\frac{f(r_0)}{r_0^{2}}-\frac{f(r)}{r^{2}}\right)
\label{drdphi2}
\end{equation}
where $r_0$ is the distance of closest approach or {\it impact parameter}. Setting $u \equiv 1/r$ in  \eqref{drdphi2} leads to
\begin{equation}
\left(\frac{du}{d\phi}\right)^2= u_{0}^2 f\left(\frac{1}{u_{0}}\right) -u^2 f\left(\frac{1}{u}\right)\label{dudphi}
\end{equation}
where $u_0=1/r_0$. Eq. \eqref{dudphi} can be integrated as
\begin{equation}
\int_{\phi_{0}}^{\phi} d\phi ' = \pm \int_{u_{0}}^{u}\frac{du'}{\sqrt{u^2_{0}f(1/u_0)-u'^2 f(1/u')}} \label{intgeod}
\end{equation}
where $\phi_0$ is the angle corresponding to the closest approach (Fig. \ref{fig:lens}). For the right part of the symmetric photon geodesic (shown in Fig. \ref{fig:lens}) $u$ decreases as $\phi$ decreases and thus we use the $+$ sign in eq. \eqref{intgeod} with $\phi_0=\frac{\pi}{2}$.

The half deflection angle $\hat \alpha/2$ at $\phi$ with respect to the straight Newtonian trajectory is the angle between the velocity vector of the photon in GR and the velocity of the photon in Newtonian theory (straight horizontal line) at each angle $\phi$. The full deflection angle requires an additional factor of 2 because the full trajectory may be thought as composed of two  parts (left and right of Fig. \ref{fig:lens}) such that both source and observer are ”far” from the lensing object \cite{Rindler:2007zz}. Thus we have
\begin{equation} 
\hat \alpha=2(\psi(\phi)-\phi)
\label{hata1}
\end{equation}
where
\begin{equation}
cos\psi=\frac{\vec r \cdot \vec v}{\vert \vec r\vert \cdot \vert \vec v \vert}
\label{cospsi}
\end{equation}
where the vectors $\vec r$ and $\vec v$ live in the $2d$ space $r-\phi$ with metric $g_{ij}=diag(f^{-1},r^2)$ with $\vec r=(r,0)$, $\vec v=(dr/d\lambda,d\phi /d\lambda)$. 
 Thus
\begin{equation}
\vec r\cdot\vec v = g_{ij}r^i v^j=f^{-1} r \frac{dr}{d\lambda}\label{rv} 
\end{equation}

\begin{equation}
\vert \vec r \vert = (g_{ij} r^i r^j)^{1/2}=f^{-1/2}r \label{mr}
\end{equation}

\begin{equation}
\vert \vec v \vert = (g_{ij} v^i v^j)^{1/2}=\left(f^{-1}\left(\frac{dr}{d\lambda}\right)^2 +r^2 \left(\frac{d\phi}{d\lambda}\right)^2 \right)^{1/2} \label{mv}
\end{equation}

Using \eqref{rv}-\eqref{mv} in \eqref{cospsi} leads to
\begin{equation}
sin\psi=\frac{f^{1/2} r}{\left(f r^2+ \left(\frac{dr}{d\phi}\right)^2\right)^{1/2}}
\label{sinpsi}
\end{equation}

Using now \eqref{drdphi2} in \eqref{sinpsi} we find
\begin{equation}
sin\psi=\frac{u(\phi)}{u_0}\frac{f(u)^{1/2}}{f(u_0)^{1/2}}
\label{sinpsiu}
\end{equation}
In what follows we use $u_0$ as a unit for $u$ and thus we set $u_0=r_0=1$ in \eqref{sinpsiu} while $u\rightarrow u/u_0$ and thus $u$ and $r$ along with all the metric parameters become dimensionless. Thus using \eqref{hata1} for the full deflection $\hat \alpha$ at angle $\phi$ and \eqref{sinpsiu} we have
\begin{equation}
\hat \alpha (\phi)=2\left[sin^{-1}\left(u(\phi)\frac{f(u)^{1/2}}{f(1)^{1/2}}\right)-\phi\right]
\label{hata2}
\end{equation}
This is a general equation for the deflection angle. Note that the effects of the metric expressed through the function $f(u)$ may be significant in metrics that are not asymptotically flat. These effects, were not taken into account in some early studies \cite{Islam:1983rxp} in the estimate of $\hat \alpha$ for a Schwarzschild-deSitter  (SdS) spacetime which is not asymptotically flat. This led to the conclusion that a cosmological constant has no effect on the gravitational lensing deflection angle. The debate on the issue of the effects of the cosmological constant on lensing however is not over \cite{Khriplovich:2008ij,Hu:2021yzn,Rindler:2007zz}. 

For a source behind the lens with respect to the observer ($\phi_{obs}\simeq 0$), an observer far from the lens ($u=O(b)<<1$), in an asymptotically flat space ($f(u)\simeq 1$) and assuming weak gravity at $r_0$ ($f(1)\simeq 1$), this equation takes the usual form \cite{Weinberg:1972kfs}
\begin{equation}
\hat \alpha \simeq 2\; u_{obs} 
\label{hata3}
\end{equation}
which is perturbatively valid to $O(b)$. This is the reason for setting $f(1)\simeq f(u)\simeq 1$ since these terms would only contribute to higher order in $b$. For asymptotically non-flat metrics however, as discussed below, we keep the contribution of $f(r)$ since for this class of metrics this term can provide $O(1)$ contribution at the large distances corresponding to $u_{obs}$. A similar approach was followed in Ref. \cite{Rindler:2007zz}.
In general, the assumption for asymptotic fltaness may not be applicable and therefore the use of \eqref{hata3} instead of \eqref{hata2} should be implemented with extreme care. This point was stressed for the first time by Rindler-Ishak (RI) \cite{Rindler:2007zz}  in the context of estimating $\hat \alpha$ in a SdS spacetime which is not asymptotically flat. In the present analysis we follow RI and impose the above assumptions $u=u_{obs}=O(b)<<1$, $f(1)\simeq 1$ (for perturbative consistency), $\phi=\phi_{obs}\simeq 0$ but we do not assume asymptotic flatness ($f(u)\simeq 1$) unless it is clearly applicable for the considered metric. Thus as shown below, our more general analysis reproduces the result of RI in the special case of SdS spacetime which is not asymptotically flat. 

In order to calculate $\hat \alpha$ under the above assumptions we thus need to obtain $u(\phi\simeq 0)\equiv u_{obs}$ as a function of the metric parameters by integrating \eqref{intgeod} with $\phi_0=\pi/2$, $\phi=\phi_{obs}\simeq 0$, $u_0=1$,  $u=u_{obs}<<1$ and with the $+$ sign and substitute it in \eqref{hata2} under the above assumptions but not assuming asymptotic flatness. Thus, the integral that needs to be valuated is
\begin{equation}
\int_{\pi/2}^{0} d\phi ' = + \int_{1}^{u_{obs}}\frac{du'}{\sqrt{1-u'^2 f(1/u')}} \label{intgeod1}
\end{equation}
Using \eqref{intgeod1}, $u_{obs}$ is expressed in terms of the metric parameters. Then, from \eqref{hata2},  $\hat \alpha$ may be obtained using
\begin{equation}
\hat \alpha \simeq 2 u_{obs}\; f(u_{obs})^{1/2}
\label{hata4}
\end{equation}
for any spherically symmetric metric of the form \eqref{sphmetric}.

We now focus on a spherically symmetric metric with
\begin{equation}
f(r)=1-b\; r^{-q}=1-b\;  u^q
\label{frt1}
\end{equation}
where $b$ and $q$ are real dimensionless parameters (rescaling with $r_0$ (or $u_0$) is assumed).
In the context of this metric, eq. \eqref{intgeod1} becomes
\begin{equation}
\int_{\pi/2}^{0} d\phi' =  \int_1^{u_{obs}}\frac{du}{\sqrt{1-u^{^2}+b(u^{^{q+2}}-1)}} \label{int1}
\end{equation}

For $b<<1$ we approximate the integral of the above equation and obtain
\begin{equation} 
-\frac{\pi}{2} = \int_1^{u_{obs}}\frac{du}{\sqrt{1-u^2}}-\frac{b}{2}\int_1^{u_{obs}} du\frac{u^{q+2}-1}{(1-u^2)^{3/2}}\label{int2}
\end{equation}
which yields
\begin{equation}
arcsin{(u_{obs})}-\frac{b}{2}\; I_q(u_{obs})\simeq u_{obs}-\frac{b}{2}\; I_q(u_{obs})=0
\label{int3}
\end{equation}
or
\begin{equation}
u_{obs}\simeq \frac{b}{2}\; I_q(u_{obs})
\label{uobsgen}
\end{equation}
where
\begin{equation}
I_q(u_{obs})=\int_1^{u_{obs}} du\frac{u^{q+2}-1}{(1-u^2)^{3/2}}
\label{iquobs}
\end{equation}
The integral $I_q(u_{obs})$ can be calculated analytically for any real $q$ as
\begin{widetext}
\ba
&&I_q(u)\equiv\int_1^u du'\frac{u'^{q+2}-1}{(1-u'^2)^{3/2}}=-\frac{u_{obs}}{\sqrt{1-{u^2_{obs}}}}-\frac{i \sqrt{\pi}\Gamma\left(-\frac{q}{2}\right)}{\Gamma\left(-\frac{1}{2}-
\frac{q}{2}\right)}-\frac{e^{-\frac{1}{2}i (3+q) \pi } \Gamma\left(-\frac{q}{2}\right) \Gamma\left(\frac{3+q}{2}\right)}{\sqrt{\pi}}\nonumber\\
&+&\frac{e^{-\frac{1}{2}i(3+q)\pi}u^{3+q}_{obs}\ _2F_1\left(\frac{3}{2},\frac{3+q}{2},\frac{5+q}{2},u^2_{obs}\right)\left(-i\cos\left(\frac{q\pi}{2}\right)+\sin\left(\frac{q\pi}{2}\right)\right)}{3+q}
\ea
\label{iqint}
\end{widetext}
The asymptotic form of this integral for $u<<1$ is
\begin{equation}
I_q(u_{obs})=\xi(q)+O(max[u_{obs},u_{obs}^{q+3}])
\label{iqser}
\end{equation}
where
\begin{equation}
\xi(q)=-\frac{i \sqrt{\pi}\Gamma\left(-\frac{q}{2}\right)}{\Gamma\left(-\frac{1}{2}-
\frac{q}{2}\right)}-\frac{e^{-\frac{1}{2}i (3+q) \pi } \Gamma\left(-\frac{q}{2}\right) \Gamma\left(\frac{3+q}{2}\right)}{\sqrt{\pi}}
\label{xiq}
\end{equation}
is a real function of $q$. In Table \ref{tabiq} we show the values of $I_q(u)$ and of $\xi(q)$ for a few integer values of $q$ obtained from the analytical expressions \eqref{iquobs}, \eqref{iqser}, \eqref{xiq}. The corresponding forms of the deflection angle $\hat \alpha$ are also shown for each $q$ using eq. \eqref{hatamq} which is derived below.

\begin{table*}
\centering
\caption{Values of $I_q(u)$ and its asymptotic form for $u<<1$ which reveals the value of $\xi(q)$ for $q\geq -2$. The corresponding forms of the deflection angle $\hat \alpha$ are also shown for each $q$ using eq. \eqref{hatamq}. The value of $q$ in each raw corresponds to the index of $I_q$. For $q\geq 0$ we have set $f(u_{obs})\simeq 1$ since $u_{obs}\ll 1$. For $q<0$ we have set $u_{obs}\simeq2m$ assuming $b\ll m$ and included the $f(u_{obs})^{1/2}$ term of eq. \eqref{hata4}. All the parameters appear in their dimensionless form where we have set the impact parameter $r_0=1$.}
\label{tabiq}
\begin{tabular}{c c c c}
\hline
$I_q$ & \textit{Analytic form} & \textit{$u<<1$ limit} & $\hat{a}(m,q) $\\[1ex] \hline\hline
\shortstack{$I_{-4}$} & $-\frac{\sqrt{1-u^2}}{u}$ & $-1/u$ & $(4m-\frac{b}{2m})(1-2m^2-\frac{b}{32m^4})$\\[1ex]
\shortstack{$I_{-3}$} & $\sqrt{\frac{2}{1+u}-1}-\tanh^{-1}(\sqrt{1-u^2})$ & $\ln{u}$ & $(4m+b\ln(2m))(1-2m^2-\frac{b}{16m^3})$ \\[1ex] 
\shortstack{$I_{-2}$} & $0$ & $0$ & $4m(1-2m^2-\frac{b}{8m^2})$\\ [1ex]
\shortstack{$I_{-1}$} & $\sqrt{\frac{2}{1+u}-1}$ & $1-u$ & $(4m+b)(1-2m^2-\frac{b}{4m})$\\[1ex]
\shortstack{$I_0$} & $\cos^{-1}u$ & $\pi/2-u$ & $4m+\frac{\pi}{2}b$\\ [1ex]
\shortstack{$I_{1}$} & $\big(u+2\big)\sqrt{\frac{2}{1+u}-1}$ & $2-u$ & $4m+2b$ \\ [1ex]
\shortstack{$I_{2}$} & $\frac{1}{2}\bigg(3\cos^{-1}u+u\sqrt{1-u^2}\bigg)$ & $\frac{3\pi}{4}-u$ & $4m+\frac{3\pi}{4}b$\\ [1ex]
\shortstack{$I_{3}$} & $\frac{-8+u\big(3+4u+u^3\big)}{3\sqrt{1-u^2}}$ & $\frac{8}{3}-u$ & $4m+\frac{8}{3}b$\\ [1ex]
\shortstack{$I_{4}$} & $\frac{1}{8}\bigg(u\sqrt{1-u^2}(7+2u^2)+15\cos^{-1}u\bigg)$ & $\frac{15\pi}{16}-u$ & $4m+\frac{15\pi}{6}b$\\[1ex] \hline
\end{tabular}
\end{table*}

For a general asymptotically flat metric ($q>0$, $f(u_{obs})\simeq 1$) we have from eqs. \eqref{hata4}, \eqref{uobsgen}
\begin{equation}
\hat \alpha_q = b\; \xi(q)
\label{hataq}
\end{equation}
For example for $q=1$ corresponding to the Schwarzschild metric we have $\xi(1)=2$ and setting $b=2m$ in \eqref{hataq} we find the well known result
\begin{equation}
\hat \alpha_{q=1}\simeq \frac{4m}{r_0}
\label{hatasch}
\end{equation}
where we have reintroduced the impact parameter $r_0$. Similarly for $q=2$ we find $I_2(u_{obs})\simeq \frac{3\pi}{4}-u_{obs}+O(u_{obs}^2) \simeq \frac{3\pi}{4}$  we find
\begin{equation}
\hat \alpha_{q=2}\simeq \frac{3\pi b}{4r_0^2}
\label{hataq2}
\end{equation}

For $q<0$ the metric \eqref{sphmetric} is not asymptotically flat and therefore the solution of \eqref{uobsgen} in general is not consistent with the assumption of $u_{obs}<<1$ while $f^{1/2}(u_{obs})$ and thus $\hat \alpha$ becomes imaginary  leading to nonphysical results.  For $q\geq -2$, $I_q(u_{obs})$ remains finite as $u_{obs}\rightarrow 0$ and is of the form $I_q(u_{obs})=\xi(q)-u_{obs}\simeq \xi(q)$ but the imaginary nature of $\hat \alpha$ remains. For example, for $q=-1$ we have $I_{-1}(u_{obs})=1-u_{obs}\simeq 1$ which leads to $u_{obs}=\frac{b}{2}$. However, from \eqref{hata4} and \eqref{uobsgen} (see also Table \ref{tabiq}) we find $\hat \alpha_{q=-1} = 2 u_{obs}\left(1-b u_{obs}^{-1}\right)^{1/2}=i \; b$ which is nonphysical. Thus even though lensing can be defined in such a spacetime, it is not consistent with the assumptions imposed above ($u_{obs}<<1$, $\phi_{obs}\simeq 0$) and thus it is beyond the scope of the present analysis.   For example, the context of our assumptions, in a deSitter spacetime ($q=-2$) there can be no lensing effect. However this conclusion is not applicable if a point mass is also present as is the case in the SdS spacetime which is not asymptotically flat. In that case even in the absence of asymptotic flatness, lensing can be well defined provided that $m>>b$\footnote{If $r_0$ gets restored this means $\frac{m}{r_0}>>b\; r_0^{-q}$.}. We will investigate this case in the next section.

\section{Lensing deflection angle in the presence of a point mass and a general fluid}
\label{IV}
The results of the previous section can be easily generalized in the case of simultaneous presence of a point mass and multiple fluids. In this case the metric function $f(r)$ takes the form
\begin{equation}
f(r)=1-\sum_{i} b_i\; r^{-q_i}=1-\sum_{i} b_i\; u^{q_i}
\label{frt2}
\end{equation}
where the sum runs over all the possible power law terms of $f(r)$ corresponding to various fluids or modified gravity. In the possible presence of a point mass in GR we can set $b_1=2m$ and $q_1=1$ allowing also for other terms. In this case, for $q_i\geq 0$, it is easy to show that the total deflection angle is obtained as a superposition of the individual deflection angles obtained from each power law term of $f(r)$ as
\begin{equation}
\hat \alpha_{tot} =\sum_i b_i \; \xi(q_i)\; f(u_{obs,i})^{1/2}\simeq \sum_i b_i\; \xi(q_i)
\label{hatasum}
\end{equation}

In the special case where one of the terms is due to a point mass ($b_1=2m$, $q_1=1$) in the context of GR and in the presence of a single additional power law term (Kiselev metric), the above equation becomes
\begin{equation}
\hat \alpha_{tot} =4m/r_0 + b\; r_0^{-q} \; \xi(q) 
\label{hatamass}
\end{equation}
where $q\geq 0$ is assumed and we have temporarily restored the unit impact parameter $r_0$ to make contact with the well known Schwarzschild deflection angle.

The case $q<0$ can also be studied provided that the additional assumption $b\ll m$ is imposed. In this case \eqref{uobsgen} gets generalized as 
\begin{equation}
u_{obs} \simeq m I_1(u_{obs}) + \frac{b}{2} I_q(u_{obs})\simeq 2m + \frac{b}{2} I_q(2m)
\label{uobsmass}
\end{equation}
since $m\gg b$. Using now \eqref{uobsmass} in \eqref{hata4} we have the general result
\begin{align}
\hat \alpha (m,q)&=(4m+b\; I_q(2m))\; f(2m)^{1/2}\nonumber\\
 &\simeq (4m+b\; I_q(2m))(1-2m^2-\frac{b}{2}(2m)^{q})
\label{hatamq}
\end{align}
This is a central result of our analysis and provides the lensing deflection angle in a general perturbed Schwarzschild metric. In the special case $q=-2$ and $b=\frac{\Lambda}{3}$ corresponding to an SdS spacetime we get from eq. \eqref{hatamq}
\begin{equation}
{\hat \alpha_{SdS}}(m,-2)\simeq \frac{4m}{r_0} (1-\frac{2m^2}{r_0^2}-\frac{\Lambda r_0^4}{24m^2})
\label{hatasds}
\end{equation}
where we have restored the unit $r_0$ for comparison with previous results. Clearly eq. \eqref{hatasds} is identical with the well known result of RI \cite{Rindler:2007zz}. 

\begin{figure}
\centering
\includegraphics[width = \columnwidth]{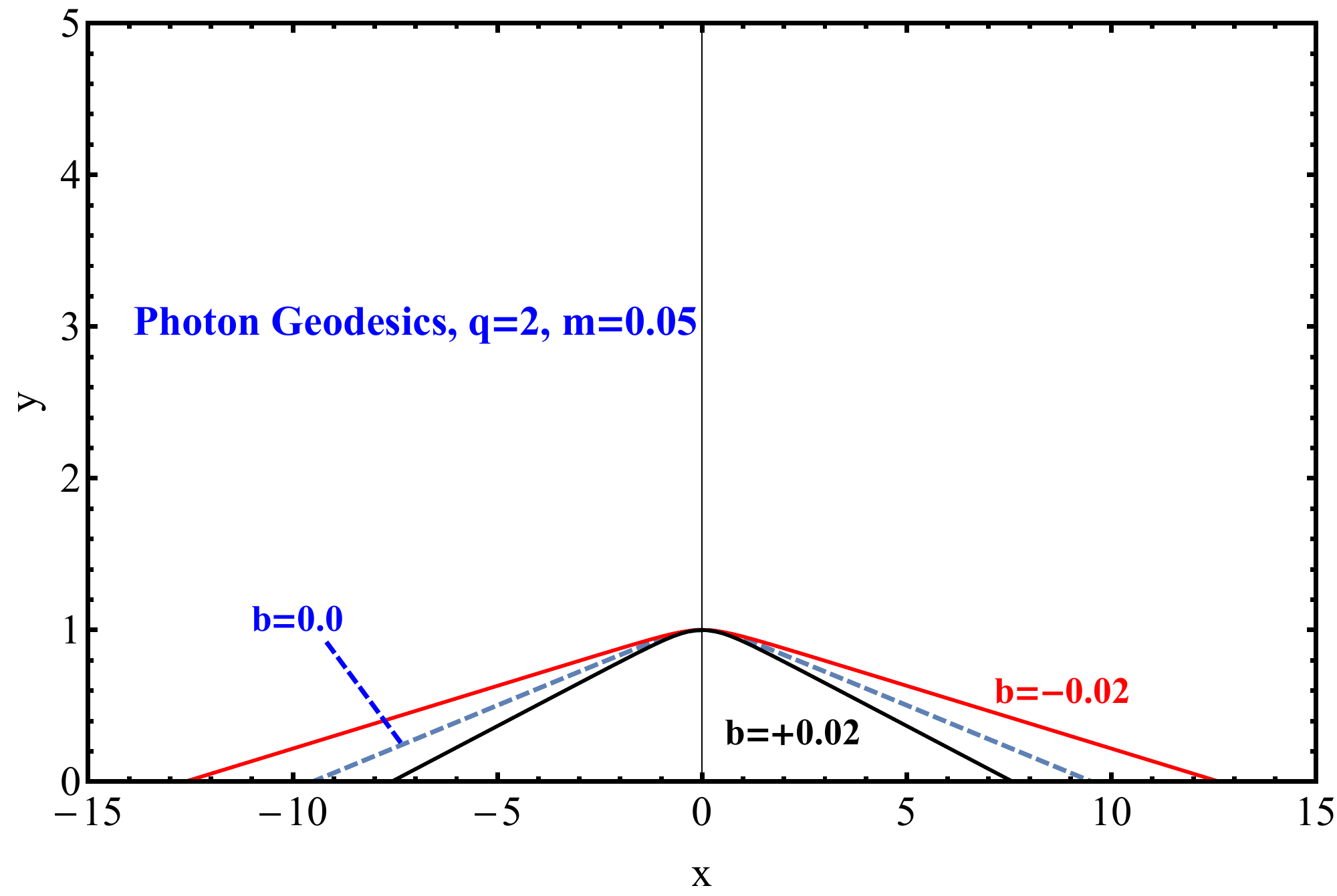}
\caption{The photon geodesic trajectories for $m=0.05$ and $b=+0.02$, $b=0$ and $b=-0.02$ with $q=2$. As expected from eq. \eqref{hatamq2}, the deflection angle decreases for $b<0$. We have set $x=r \cos\phi$ and $y=r\sin \phi$.}
\label{fig3}
\end{figure}

Similarly, for $q=2$, eq. \eqref{hatamq} reduces to 
\begin{align}
\hat \alpha(m,2)&=(4m+b\; I_2(2m))(1-2m^2-\frac{b}{2}(2m)^2)\nonumber\\&\simeq 4m +b\; \xi(2)= \frac{4m}{r_0}+\frac{3\pi b}{4r_0^2}
\label{hatamq2}
\end{align}
where in the last equality we have restored the unit impact parameter $r_0$. This result is also consistent with the corresponding result of the previous section \eqref{hataq2}. It is therefore clear that eq. \eqref{hatamq} provides a general result that generalizes the corresponding result of RI \cite{Rindler:2007zz} applicable in a wide range of spherically symmetric metrics. In Fiq. \ref{fig3} we show the photon geodesics for $m=0.05$, $q=2$ for three values of the parameter $b$. As expected from eq. \eqref{hatamq2}, the deflection angle increases with $b$.

\begin{figure}
\centering
\includegraphics[width = \columnwidth]{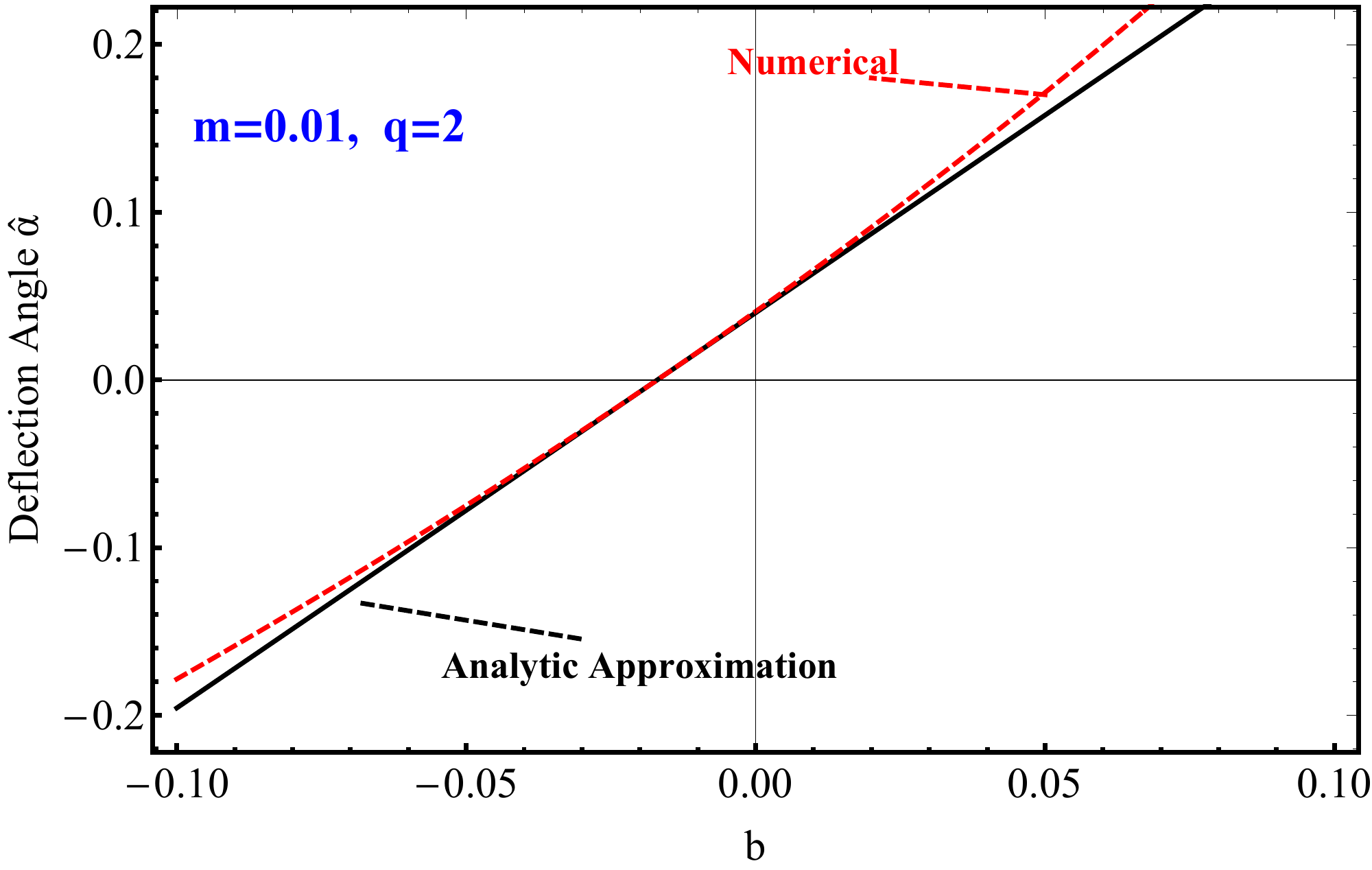}
\caption{A comparison of the exact numerical result for the deflection angle with the approximate analytic result of \eqref{hataq2} for $m=0.01$, $q=2$. The numerical result was obtained by solving numerically eq. \eqref{intgeod1} for $u_{obs}$ and then using \eqref{hata2} with $\phi=0$. For small $b$ and $\hat \alpha$ the agreement is very good. Note that the assumption $m>>b$ is not needed for $q>0$.}
\label{fig2}
\end{figure}

 In order to test the validity of the above analytical results we have compared them with exact numerical solution for the deflection angle obtained  by solving numerically eq. \eqref{intgeod1} for $u_{obs}$ and then using \eqref{hata2} with $\phi=0$ to obtain $\hat \alpha$ for fixed values of $m$ and $q$. The comparison of the approximate analytic result of \eqref{hataq2} for $m=0.01$, $q=2$ with the corresponding exact numerical result is shown in Fig. \ref{fig2}. The corresponding photon geodesic trajectories for $m=0.01$ and $b=+0.02$, $b=0$ and $b=-0.02$ with $q=2$ are shown in Fig. \ref{fig2}. As expected from eq. \eqref{hatamq2}, the deflection angle decreases for $b<0$. For illustration purposes we plotted the photon geodesics for $q=2$. However, our results and in particular eq. (\ref{hatamq}) are general and valid for any value of the parameter $q$.

Eq. \eqref{hatamq} can be used to obtain observational constraints on the parameters $m,b,q$ from cluster Einstein radius lensing data thus constraining the possible presence of exotic fluids and/or modified gravity in cluster dynamics. A method for obtaining such constraints is illustrated in the next section.

\section{Observational Constraints on the Metric Parameters}
\label{V}

\begin{figure}[h]
\centering
\includegraphics[width = \columnwidth]{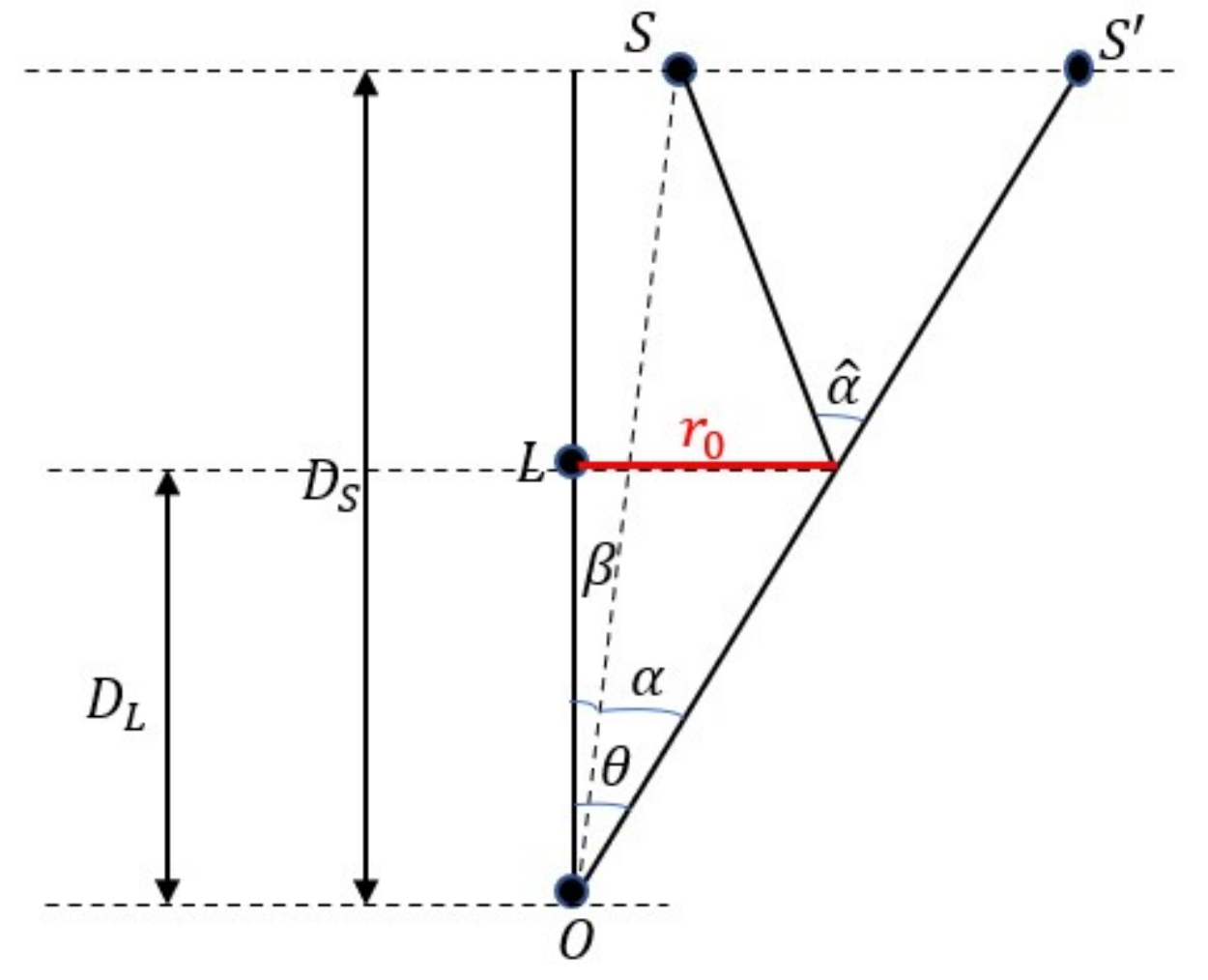}
\caption{The geometry of a lensing event and the definition of the relevant angles and distances.}
\label{fig:figlens}
\end{figure}

In order to obtain an order of magnitude estimate of the upper bound of the parameter $b$ of eq. \eqref{hatamass} we consider a sample of clusters \cite{Cutajar:2014gfa} which act as lenses to background galactic sources. In Fig. \ref{fig:figlens} we show the typical lensing diagram where in our analysis we have assumed $\beta \simeq 0$. 

The deflection angle is not directly observable but rather inferred from the observed positions of the images in the sky after gravitational lensing. Without knowledge of the true source position, the distances to the lens and source, and other constraints, the deflection angle can not be directly estimated. We thus rely on lens modeling techniques that use observed image positions, lens and source distances, as well as additional data on redshift to infer the deflection angle and other lens parameters.

Let $D_L$ be the distance to the lens (cluster) at redshift $z_L$ and $D_S$ the distance to the background lensed galaxy at redshift $z_{arc}$ while the observed Einstein ring of the source appears at angle $\theta$. The distances to the lens and the source may be approximately obtained from the corresponding redshift using the angular diameter distance of the form
\begin{equation}
D(z)=\frac{c\; z}{H_0}+\frac{c(1-q_0)}{2H_0} z^2 +O(z^3)
\label{distz}
\end{equation}
where we have restored the speed of light $c$ (set to 1 in previous sections). In what follows we assume $H_0=70 km/(sec \dot Mpc)$ and $q_0 = -0.5$. Therefore, $z_L$ (and thus $D_L$), $z_S$ (and thus $D_S$) and the Einstein angle $\theta$ are measurable.  Then, the deflection angle $\hat \alpha$ can be obtained using the measured quantities. In particular from Fig. \ref{fig:figlens} we have
\begin{equation}
r_0 \simeq D_L \; \theta \label{r0dist}
\end{equation}
\begin{equation}
\hat \alpha \; (D_S-D_L) \simeq \theta \; D_S \label{hatadist}
\end{equation}
Eq. \eqref{hatadist} can be used for the measurement of the deflection angle induced by cluster lenses. Through such a measurement constraints on the metric parameters can be imposed.  

For example if the deflection is assumed to be induced by the cluster mass only, eq. \eqref{hatadist} can lead to constraints on the cluster mass $M_{cl}$. In particular, for the cluster A2218, using the entries of Table \ref{tabdata} and restoring $G$ and $c$ in eq. \eqref{hatasch} we have
\begin{equation}
\hat \alpha = \frac{4\; G \; M_{cl}}{c^2 r_0}=\frac{4\; G\; M_{cl}}{c^2 \theta D_L}=\theta \frac{D_S}{D_S-D_L}\simeq 10^{-4}
\label{mfrhata}
\end{equation}
which leads to $M_{cl}\lesssim 5 \times 10^{13} M_\odot$. 

Similarly, the last column of Table \ref{tabdata} may be used to derive constraints on the metric coefficients $b_q$ for any value of the power coefficient $q$. This  column provides an estimate of the product $b_q r_0^q$ if the full lensing was due to this metric term. Since, at least a large part of the lensing is due to a Schwarzschild term of the metric, the values of this column may be viewed as upper bounds of the product $b_q r_0^q$ where $r_0$ is the impact parameter which is specified for each system. Thus, once $q$ is specified, it is straightforward to use the data in that column to obtain upper bounds on the metric coefficients $b_q$. This is demonstrated below  where an upper bound on the metric coefficient $b_q$ is obtained.

Thus, if we assume that a $q$-exotic fluid is solely responsible for the lensing we have
\begin{equation}
\hat \alpha = b_q I_q r_0^{-q} \simeq \theta \frac{D_S}{D_S-D_L}\simeq 10^{-4} 
\label{besthata}
\end{equation}
which leads to an order of magnitude estimate $b_q r_0^{-q} \lesssim 10^{-4}$ or $b_q \lesssim 10^{-4} \theta D_L\simeq 10^{-4-q} Mpc^q$.

In a similar way we may obtain order of magnitude constraints from all clusters shown in Table \ref{tabdata}. Such constraints are shown in the last column of Table \ref{tabdata}.

\begin{table*}
\centering
 \caption {Order of magnitude constrains on the metric parameters $b_q$ from cluster lensing data.}
\label{tabdata}
\begin{tabular}{c c c c c c c}
\hline
$\text{Cluster}$  & \shortstack{$\theta (arcsec)$} & \shortstack{$z_L$} & \shortstack{$z_{arc}$} & \shortstack{$\hat{a}(arcsec)$} & \shortstack{$r_0\simeq\theta D_L(Mpc)$} & \shortstack{$b_q r_0^{-q}(\cdot 10^{-4})$} \\ [1ex] \hline\hline
$\text{A}370$ & $39.0$ & $0.375$ & $0.725$ & $80.4$ & $0.304$ & $3.9$\\[1ex] 
$\text{A}370$ & $45.2$ & $0.373$ & $1.3$ & $63.9$ & $0.350$ & $3.1$\\ [1ex] 
$\text{A}963$ & $18.7$ & $0.206$ & $0.711$ & $26.8$ & $0.080$ & $1.3$ \\[1ex] 
$\text{A}1689$ & $45.0$ & $0.183$ & $1.000$ & $55.7$ &$0.171$ & $2.7$ \\[1ex] 
$\text{A}2163$ & $15.6$ & $0.201$ & $0.728$ & $22.7$ & $0.065$ & $1.1$ \\[1ex] 
$\text{A}2218$ & $20.8$ & $0.175$ & $0.702$ & $26.7$ & $0.076$ & $1.3$ \\[1ex] 
$\text{A}2218$ & $23.7$ & $0.171$ & $0.515$ & $35.1$ & $0.084$ & $1.7$ \\[1ex]  
$\text{A}2218$ & $73.2''$ & $0.171$ & $1.034$ & $88.7$ & $0.260$ & $4.3$ \\[1ex]  
$\text{A}2390$ & $37.4$ & $0.228$ & $0.913$ & $49.5$ & $0.177$ & $2.4$ \\[1ex] 
$\text{MS}0440$ & $21.7$ & $0.197$ & $0.530$ & $35.1$ & $0.089$ & $1.7$ \\ [1ex] 
$\text{MS}1358$ & $17.7$ & $0.329$ & $4.92$ & $18.6$ & $0.121$ & $0.9$ \\[1ex] 
$\text{PKS}0745$ & $21.4$ & $0.103$ & $0.433$ & $28.9$ & $0.046$ & $1.4$ \\ [1ex] \hline
\end{tabular}
\end{table*}

\section{Conclusion-Discussion}
\label{VI}

We have derived an analytic expression that provides the lensing deflection angle in a generalized spherically symmetric metric. This result extends previous special cases of spherically symmetric metrics including the Schwarzschild,  SdS and spherical Rindler metrics. Our results have been tested using exact numerical solutions and reduce to previously known results in special cases of perturbed Schwarzschild metrics. Using the Einstein radii around clusters we have imposed order of magnitude constraints on the new parameters of the metric. Our results are valid to first post-Newtonian order but the method may be generalized to higher nonlinear orders by including more terms in the expansion of the integral \eqref{int1}.

Our results can be useful in the context of recent analyses on gravitational lensing in modified gravity theory (eg \cite{Islam:2020xmy, Poshteh:2018wqy, Kuang:2022ojj, Kumar:2021cyl}. These and other corresponding analyses involve generalized spherically symmetric metrics which in most cases reduce to our general metric described by eqs. (\eqref{sphmetric}, \eqref{fgen1}, \eqref{grser}
) at large distances where specific power laws dominate. Thus, our analysis can be used as a limiting testing case for most generalized modified gravity spherically symmetric metrics where the modified gravity degrees of freedom allow deviations from the standard Schwarzschild metric of GR.

The exotic fluids we have considered could be manifestations of either dark matter or dark energy in clusters of galaxies. Therefore an interesting extension of the present analysis would be to compare the derived constraints with dark energy constraints obtained using probes of the cosmic expansion rate like standard rulers (eg the CMB sound horizon) or standard candles (eg SnIa). 

The generalized spherically symmetric metric considered here could emerge as vacuum solutions in modified gravity models like the Grumiller metric \cite{Grumiller:2010bz} or Weyl vacuum \cite{Mannheim:1988dj}. In that case the imposed constraints on the metric parameters may be translated to constraints of the corresponding modified gravity Lagrangian parameters. This would also be an interesting extension of the present analysis. Similarly, a generalization of the axisymmetric Kerr metric with dark fluid power law terms and the derivation of the corresponding lensing deflection angle would constitute an interesting extension of the present analysis \cite{Ghosh:2022mka,Molla:2021sgw,Bozza:2002af}. 

Finally, the consideration of other general spherically symmetric spacetimes \cite{Mantica:2022phm, Gurtug:2020kwd} and the derivation of the corresponding deflection angle in terms of the metric parameters could also lead to a generalization of the present analysis.

\section*{Acknowledgments}
This article is based upon work from COST Action CA21136 - Addressing observational tensions in cosmology with systematics and fundamental physics (CosmoVerse), supported by COST (European Cooperation in Science and Technology). This project was also supported by the Hellenic Foundation for Research and Innovation (H.F.R.I.), under the "First call for H.F.R.I. Research Projects to support Faculty members and Researchers and the procurement of high-cost research equipment Grant" (Project Number: 789).

\section*{Data Availability Statement}
The Mathematica (v12) files used for the production of the figures and for derivation of the main results of the analysis can be found at \href{https://github.com/leandros11/lensing1}{this Github repository under the MIT license.} \\


\raggedleft
\bibliographystyle{mnras}
\bibliography{main} 



\bsp	
\label{lastpage}
\end{document}